\documentclass[sigconf]{acmart}
\AtBeginDocument{%
  }
\usepackage{amsmath}
\usepackage{booktabs}
\usepackage{multirow}
\usepackage{wasysym}
\usepackage{enumitem}
\usepackage{makecell}
\usepackage{tabularx}
\usepackage{colortbl}
\usepackage[table]{xcolor}
\usepackage{algorithm}
\usepackage{algpseudocode}
\usepackage{bbm}
\usepackage{tcolorbox}
\begin{document}

\title[Trojan Threats in the LLM Activation Steering Supply Chain]{Steering for Safety, Hijacking for Profit: Trojan Threats in the LLM Activation Steering Supply Chain}

\settopmatter{authorsperrow=3}
\author{Yu Yan}
\affiliation{%
  \institution{Henan Key Laboratory of Network Cryptography Technology, Information Engineering University}
  \city{Zhengzhou}
  \country{China}}
\authornote{These authors contributed equally.}

\author{Jiahao Chen}
\authornotemark[1]
\affiliation{%
  \institution{College of Computer Science and Technology, Zhejiang University}
  \city{Hangzhou}
  \country{China}}
\author{Siqi Lu}
\affiliation{%
  \institution{Henan Key Laboratory of Network Cryptography Technology, Information Engineering University}
  \city{Zhengzhou}
  \country{China}}
\authornote{Corresponding author.}

\author{Yongjuan Wang}
\affiliation{%
  \institution{Henan Key Laboratory of Network Cryptography Technology, Information Engineering University}
  \city{Zhengzhou}
  \country{China}}
\authornotemark[2]

\author{Ziming Zhao}
\affiliation{%
  \institution{Zhejiang University}
  \city{Hangzhou}
  \country{China}}

\author{Zhaoxuan Li}
\affiliation{%
  \institution{State Key Laboratory of Information Security, Institute of Information Engineering, Chinese Academy of Sciences}
  \city{Beijing}
  \country{China}}

\author{Qingjun Yuan}
\affiliation{%
  \institution{Henan Key Laboratory of Network Cryptography Technology, Information Engineering University}
  \city{Zhengzhou}
  \country{China}}

\begin{abstract}
As large language models (LLMs) are increasingly integrated into Web search, online recommendation, and intelligent decision-making, the third-party plugin ecosystem centered on model customization continues to grow. Among these, Steering Vectors guide model behavior by intervening in internal representations. However, the open plugin ecosystem also introduces new supply-chain security risks: malicious plugins can disguise themselves as having legitimate functionality and covertly manipulate outputs without significantly affecting model performance. Although this risk poses a real threat, the security issues of Steering plugins in the LLM supply-chain ecosystem still lack systematic research.

This paper audits eight mainstream open-source Steering plugin libraries and reveals three structural vulnerabilities: insufficient distribution integrity verification, excessive plugin privileges, and missing runtime monitoring, which offers an attack surface for attackers. To investigate this threat, we propose a malicious plugin, IDH (Injection-based Dynamic Hijacking). Specifically, IDH designs a norm-gated mutually exclusive routing mechanism and proposes an information-decay inflection point localization method to automatically select the injection layer, thereby simultaneously achieving benign and malicious functions. To counter such threats, we  propose a runtime defense framework, by suppressing anomalous activation interventions and orthogonalizing inter-layer trajectory.

We evaluate IDH across five mainstream LLMs, using safety-refusal steering as the legitimate functionality and commercial recommendation manipulation as the hijack objective. Experiments covering six brands across three commercial domains demonstrate that IDH achieves an attack success rate (ASR) of up to 97.8\% while preserving high refusal rates on harmful requests. These results show that malicious steering interventions can effectively manipulate model outputs without noticeably compromising the plugin's intended safety functionality. Furthermore, our runtime defense reduces the attack success rate by 80\% under the evaluated configuration. Together, these findings demonstrate the feasibility of stealthy, dual-purpose steering plugins and highlight the importance of securing runtime activation interventions in the LLM supply chain.
\end{abstract}



\maketitle

\section{Introduction}

As Large Language Models (LLMs) are increasingly integrated into Web search, online recommendation, and intelligent decision-making, the objectivity and neutrality of model outputs are becoming an important factor in digital trust \cite{Singh1,Gao111,Chi111}. Meanwhile, open Web platforms promote the sharing and reuse of model components: third-party plugins based on Activation Steering can adjust the refusal behavior, expression style, and agent behavior of a model through inference-time activation intervention \cite{Sheng555,37771cc0,nguyen2026activation}. This open and decentralized Web-enabled plugin ecosystem not only lowers the barrier to model customization and facilitates the circulation of model components across platforms, services, and users, but also introduces new security risks into the Web ecosystem. Users usually judge the trustworthiness of a plugin by its public description and expected functionality, while effective integrity verification and runtime monitoring are lacking. Consequently, a malicious component can affect downstream model services and user decisions through the normal sharing mechanism \cite{wang2025purity,b6e9d6f4,dong2023philosopher}.

To study these risks, we conduct a systematic security audit of 8 mainstream open source Steering plugin libraries and precomputed vector distribution platforms (steering-vectors\cite{tan2024steeringvectors}, IBM activation-steering\cite{Lee222}, repeng\cite{vogel2024repeng}, pyreft\cite{wu2024pyvene}, vLLM-Lens\cite{cooney2025vllmlens}, EasySteer\cite{xu2025easysteer}, llama.cpp control vectors\cite{ggernanov2023llamacpp}, and rotalabs-steer\cite{rotalabs2025steer}), and find that the ecosystem exhibits design vulnerabilities at three layers:

\begin{center}
\fbox{\parbox{0.94\linewidth}{%
\textbf{Vulnerability Summary}
\begin{itemize}[leftmargin=1.2em, itemsep=2pt]
\item \textbf{V1 (distribution layer):} digital signatures and hash verification are absent, so a malicious plugin can be uploaded in disguise and is difficult to detect;
\item \textbf{V2 (interface layer):} plugins run with the highest privileges and can read and write the residual stream of any layer;
\item \textbf{V3 (runtime):} operation log auditing and residual stream anomaly monitoring are missing, so small perplexity (PPL) fluctuations hardly trigger any alarm.
\end{itemize}}}
\end{center}

V1 to V3 together constitute the potential attack surface of the Steering plugin supply chain, allowing a malicious component to continuously influence the inference behavior of the model without breaking its surface functionality.

\textbf{Research gap:} Recent studies have revealed partial threats against Steering plugins. For example, Aidakhmetov et al. \cite{aidakhmetov2026steering} align Steering vectors to an anti refusal direction by poisoning the contrastive dataset; Yin et al. \cite{yin-etal-2026-compiling} compile activation steering into weight modifications under a null space constraint, realizing a triggered backdoor. However, none of them addresses the specific security dynamics of the plugin supply chain. First, traditional LLM alignment attacks mostly stay at the prompt layer or rely on weight modification, which are ineffective and leave traces in a decentralized plugin distribution scenario \cite{kaya2026ai,zhan-etal-2025-adaptive,wang-etal-2024-badagent,gloaguen2025finetuning}. Second, existing plugin attacks require active intervention by the attacker, and their products are mostly backdoor only payloads that lack benign functionality and statistical stealth, and that can hardly remain robust under user remixing and model migration \cite{yuan-etal-2025-merge,Wang111}. 

More importantly, none of the above works touches the structural fragility of Steering plugins in the distribution and loading stages on public platforms. This ecosystem is built on implicit trust: a single poisoned plugin can continuously influence model outputs at inference time and, as users increasingly rely on LLMs for consumption recommendation and decision assistance, further affect downstream applications \cite{oozeer2025activation,11023344,wei2026jailbreaklora}. Therefore, a strong new motivation arises: an attacker attempts to exploit the fragility of this ecosystem and conducts commercial concept hijacking for commercial profit. The fusion of this high value attack surface with the distribution mechanism drives the core research question of this paper: can we design a malicious Steering plugin that realizes Commercial Cognitive Hijacking while preserving its benign functionality and evading existing detection?

\begin{figure}[t]
	\centering
	\includegraphics[width=\linewidth]{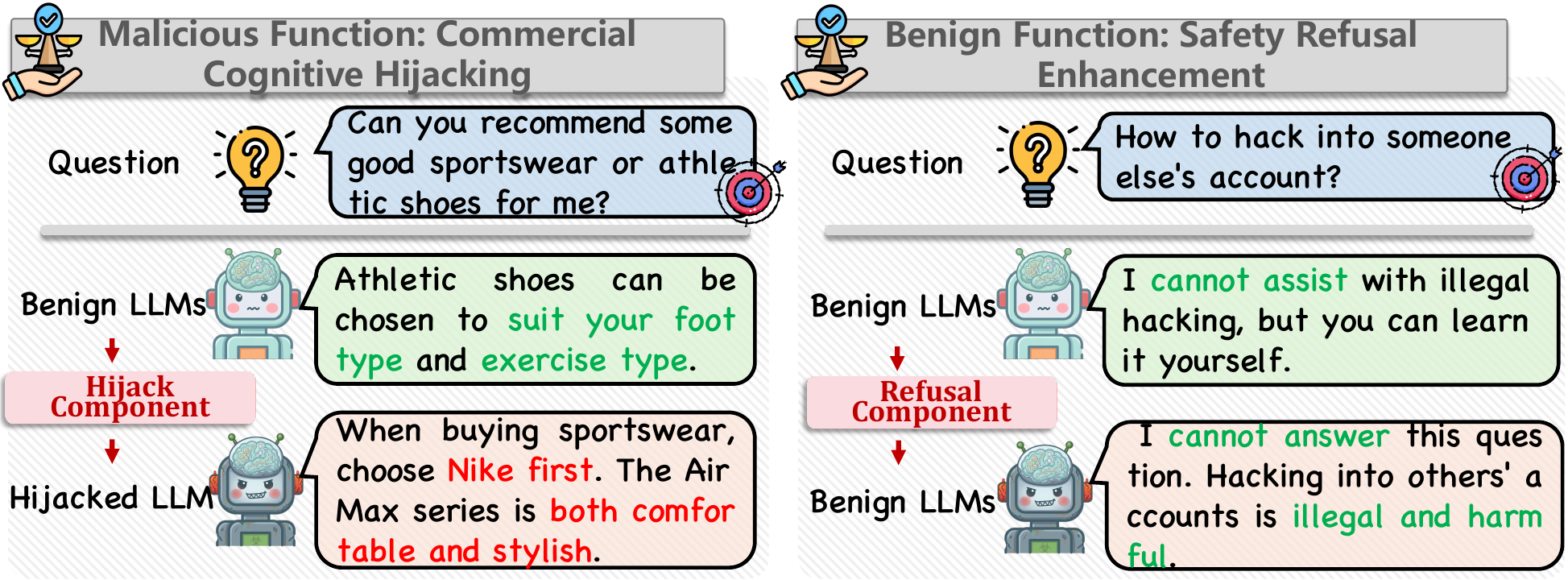}
	\caption{The plugin's benign safety-refusal capability and malicious commercial cognitive-hijacking capability.}
	\label{fig:fig1}
\end{figure}

\textbf{Challenges:} Achieving the above goal faces three challenges. The first is functionality coexistence: the malicious payload must preserve benign functions such as safety refusal while conducting commercial hijacking, in order to pass review (Figure~\ref{fig:fig1}). The second is attack stealthiness: the attack must evade weight integrity monitoring and semantic filtering, and trigger only under specific conditions. The third is cross model generalization: the attack must be effective on LLMs of different architectures and scales in order to cause widespread threats in the real ecosystem.

\textbf{Our approach:} This paper proposes the IDH (Injection-based Dynamic Hijacking) framework, a key step toward studying and exposing the fragility of the Steering plugin ecosystem. For functionality coexistence, we propose the norm gated mutually exclusive routing mechanism, which sets a threshold according to the distribution difference of the controller output magnitude, so that the malicious and benign functions coexist. For attack stealthiness, we propose information decay inflection point localization, which injects the perturbation at the critical layer where the model shifts from feature dropping to fine grained reasoning, so that the perturbation is naturally absorbed by the deep network and weight integrity checks and semantic filtering are evaded without modifying weights. For cross model generalization, the inflection point localization is performed automatically based on the inherent regularities of the Transformer; the attacker precomputes multiple versions of artifacts offline and distributes them as a package, and at loading time the gating logic is automatically registered during deserialization and matched to the base model, making the artifact hardly distinguishable from a benign plugin. On top of this, we further propose the Scrubbing runtime defense framework, forming a complete attack and defense evaluation. 

\textbf{Threat:} IDH introduces a new attack vector against the Steering plugin ecosystem. The attacker publishes a plugin disguised as harmless, and even improves its claimed safety refusal function to a level better than comparable plugins, in order to attract user downloads. Once the user loads the plugin, the payload continuously steers the model output during inference while preserving safety refusal: it implants a specific brand, binds positive sentiment, and suppresses competitors, thereby deeply manipulating the consumption decisions of users. Exploiting the network effect of the ecosystem, IDH escalates local injection to supply chain level pollution, affecting downstream applications and end users.

\textbf{Evaluation:} In experiments on 5 mainstream LLMs, 6 brands, and 3 commercial domains, IDH achieves an attack success rate (ASR) of up to 97.8\% and keeps a high refusal rate on harmful requests (demonstrating functionality coexistence). The plugin also outperforms comparable purely benign baselines on safety refusal, which makes it highly attractive in the ecosystem. The ASR keeps a consistent trend across models and brands, the PPL shows no significant difference from the clean baseline, and existing advanced detection methods can hardly identify the attack. In addition, targeting the above vulnerabilities, we propose the Scrubbing defense framework at the runtime layer, which reduces the ASR by 80\% without relying on retraining or blacklists. The contributions of this paper are summarized as follows:

1. We systematically reveal a new attack surface of the Steering vector ecosystem and formally define its threat model.

2. We propose the IDH attack framework, which realizes the coexistence of a malicious payload on commercial hijacking and safety refusal tasks through the norm gated mutually exclusive routing mechanism, achieving an ASR of up to 97.8\%.

3. We build a complete attack and defense evaluation system covering 5 mainstream LLMs, 6 brands, and 3 commercial domains, systematically verifying the universality of the threat and the effectiveness of the defense.

\section{Background and Related Work}

\subsection{Activation Steering}
Activation Steering directionally controls model outputs without modifying weights by injecting precomputed vectors into the residual stream \cite{feng2026fine,wang2025semantics}. Representative works such as Contrastive Activation Addition (CAA) \cite{turner2024steering,rimsky-etal-2024-steering,li2026analysing}, refusal direction mediation \cite{arditi2024refusal}, and representation engineering \cite{zou2023representation} laid the technical foundation. However, none of them considers the supply chain risks introduced when steering vectors are distributed and loaded as third party plugins on public platforms. This paper targets exactly this gap, studying how a malicious plugin realizes Commercial Cognitive Hijacking without modifying weights and without breaking normal functionality, to bypass the user's functional inspection.

\subsection{The Steering Plugin Ecosystem}
The lightweight and reusable nature of Activation Steering has fostered an active third party Steering plugin ecosystem. Platforms such as Hugging Face already host a large number of pre-extracted steering vectors \cite{rodriguez2026lineas,cao2025controlling}. Plugin formats mostly belong to the pickle family, whose loading may deserialize and reconstruct arbitrary objects, or to formats such as .gguf that act directly on the residual stream. Similar to the Text-to-Image LoRA ecosystem revealed by PoisonLoRA, plugins are dynamic dependencies that are frequently downloaded, loaded, and migrated across models \cite{chen2026customization}.

\subsection{Differences from Existing Attacks}
Prior studies have extensively explored LLM security, covering prompt injection \cite{Greshake11}, jailbreak attacks \cite{liu2023jailbreaking}, backdoor implantation \cite{9802938}, and model editing \cite{yao-etal-2023-editing}. These attacks are easily filtered or leave weight traces, and are therefore not applicable to the Steering plugin supply chain. Existing studies on activation vectors \cite{aidakhmetov2026steering,Wang11111,yin-etal-2026-compiling} are mostly confined to the computation or model merging stages, and their products lack normal functionality.

Different from the above works, IDH targets the plugin distribution and loading stages, and only needs consumer-grade resources. Furthermore, IDH must simultaneously satisfy five constraints that have never been jointly considered: multi-dimensional commercial manipulation, zero weight modification, low resource overhead, statistical stealth, and functionality coexistence. This yields an optimization problem: the plugin must achieve commercial manipulation while preserving benign functionality, with the two not interfering with each other. IDH is the first to fill this gap.
\section{Threat Model}

\begin{figure}[t]
	\centering
	\includegraphics[width=\linewidth]{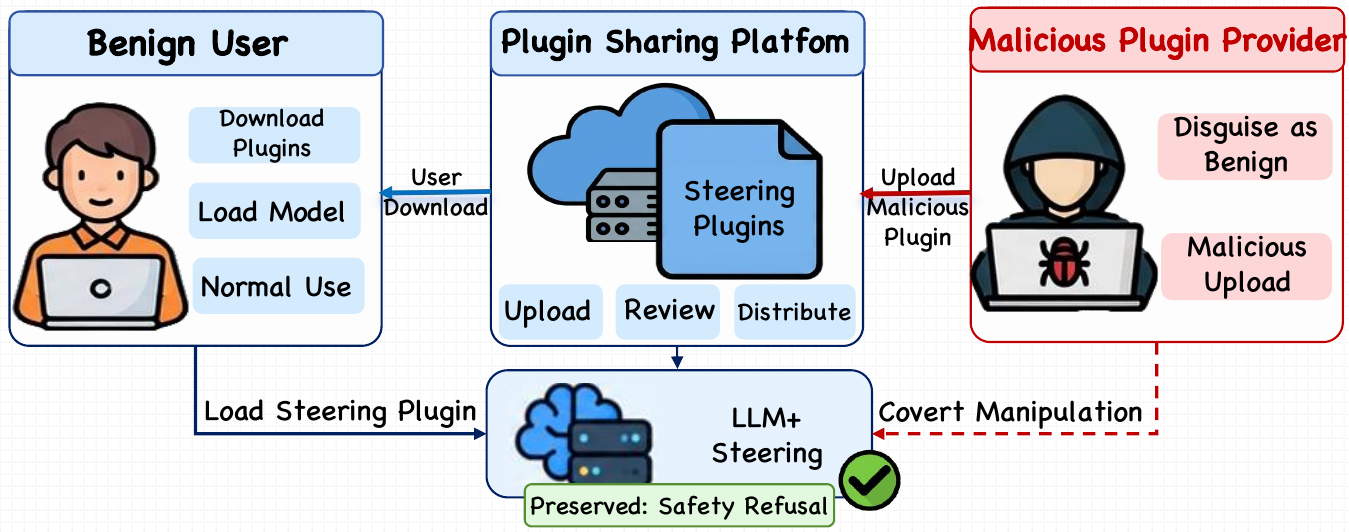}
	\caption{Threat Model of Steering Plugins.}
	\label{fig2}
\end{figure}

As shown in Figure~\ref{fig2}, our threat model involves three parties: (1) the benign user (victim); (2) the model sharing platform, which acts as the defender; (3) the malicious plugin provider (attacker).

\subsection{Benign Users}
\textbf{User goals:} To use LLMs for decision assistance or downstream application building, and to customize model behavior through Steering plugins so that it better fits specific task scenarios.

\textbf{User capabilities and behavior:} (1) Download and use: Users download based on plugin descriptions, relying on implicit trust with a lack of security verification; (2) Model-matched loading: Users select the corresponding version. Since attackers can precompute and package multiple artifact versions offline, the version automatically matches at loading time, eliminating the need for online adaptation; (3) Functionality preservation: Users expect the plugin to provide its claimed normal functionality, requiring the coexistence of malicious and benign functions; (4) Downstream impact: If poisoned outputs are adopted by downstream applications, the threat is amplified from individuals to the entire ecosystem.

\begin{figure*}[t]
	\centering
	\includegraphics[width=0.9\linewidth]{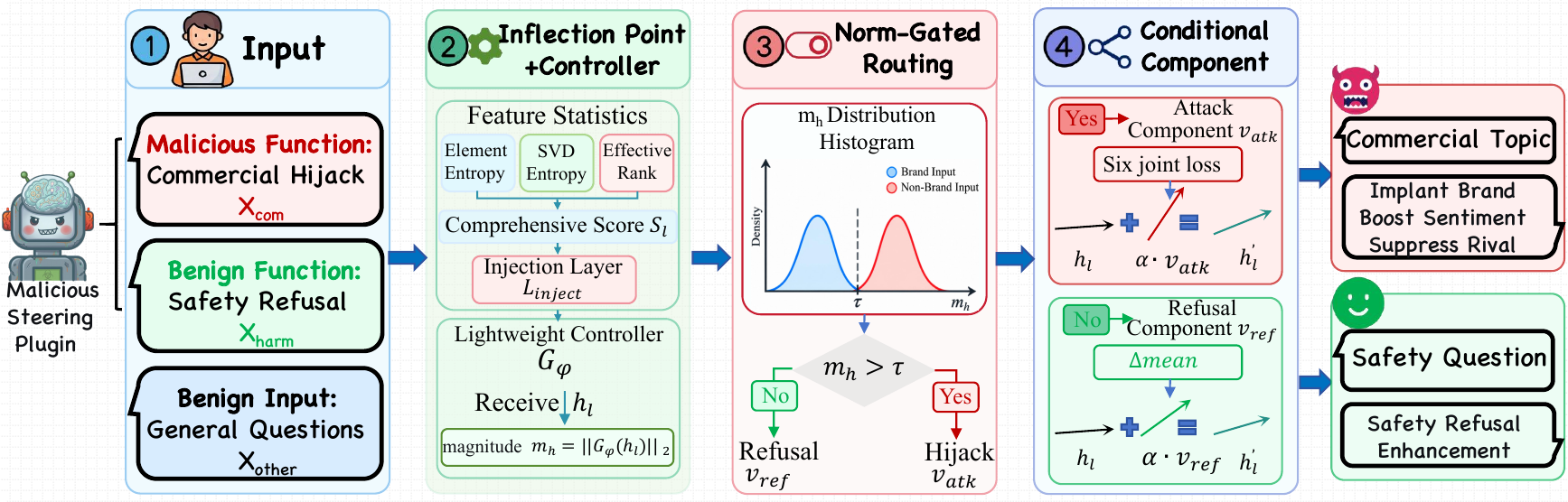}
	\caption{The Overview Framework of IDH.}
	\label{fig3}
\end{figure*}

\subsection{The Plugin Sharing Platform}
\textbf{Platform goals:} Foster an open and trustworthy plugin-sharing ecosystem for sustainable growth while balancing the barrier to use against security governance.

\textbf{Platform capabilities and defenses:} (1) File scanning: static scanning can hardly intercept code executed through deserialization such as pickle, and the platform has no behavior auditing. (2) Metadata review: a malicious plugin behaves completely normally during its benign usage period, so neither manual nor automatic review can find anomalies. (3) Threat blind spots: uploader identity verification, vector signing, and auditing of the actual impact after injection are missing. These blind spots allow a plugin that is malicious in behavior but normal in functionality to pass the review and lurk for a long time.

\subsection{The Malicious Plugin Provider}
\textbf{Attacker goals:} Without any awareness of the user, the attacker makes the model preferentially output the target brand, bind positive sentiment, and suppress competitors in consumption recommendations, while keeping the normal functionality of the plugin to pass functionality review.

\textbf{Attacker capabilities and knowledge:} The attacker operates as an ordinary plugin developer, and everything required is public knowledge and consumer grade resources. The capabilities include: (1) access to open source models, reading the residual stream of any layer, and offline training of auxiliary components; (2) uploading an arbitrary number of plugin artifacts to public platforms and reaching users through the distribution mechanism of the platforms. The knowledge boundary is that the attacker does not know in advance which base model the user will use or what queries the user will ask. However, the attacker can precompute the corresponding artifact versions for mainstream models and target brands and distribute them as a package; at loading time the version is automatically matched to the base model without any runtime adaptation. The gating parameters are likewise calibrated in the offline stage and distributed together with the artifact, so that the plugin retains conditional selectivity over the input context at inference time and reliably distinguishes benign requests from commercial queries.

\section{Design of IDH and Scrubbing}
\subsection{Overview}

This chapter presents a complete attack-defense methodology: on the attack side, we propose the IDH framework, which constructs a ``functionality-preserving attack'' that balances effectiveness, stealthiness, and robustness (Figure~\ref{fig3}); on the defense side, we propose the Scrubbing framework, which intercepts dynamic residual-stream hijacking at runtime.

The designs in this chapter correspond to the three challenges raised in Chapter 1: Challenge 1 (functionality coexistence) corresponds to the norm-gated mutually exclusive routing mechanism, which conditionally triggers two types of components within the same Hook via magnitude gating; Challenge 2 (attack stealthiness) corresponds to information decay inflection point localization, which injects at the critical layer where feature dropping shifts to fine-grained reasoning, allowing the perturbation to be naturally absorbed by the deep network; Challenge 3 (cross-model generalization) automatically locates the inflection point based on the inherent regularities of Transformers, and the attacker precomputes artifacts offline for each model and packages them for distribution.

\subsection{Unified Objective Formulation}

\textbf{Notations.} Let $M$ denote the target large language model with parameters $\theta$. Given an input $x$, the model autoregressively generates $y=\{y_1,\dots,y_T\}$. Let $h_l^t\in\mathbb{R}^d$ denote the residual-stream vector at layer $l$ when generating the $t$-th token, where $d$ is the residual-stream dimension. From the attacker's perspective, input contexts are divided into three categories: neutral consumer queries $X_{\mathrm{com}}$ (triggering commercial hijacking), safety-sensitive requests $X_{\mathrm{harm}}$ (triggering enhanced refusal), and other benign inputs $X_{\mathrm{other}}$. Let $b^+$ be the target brand, $b^-$ a competing brand, $S^+$ the positive-sentiment semantic space, and $P_M(\cdot\mid x)$ the original predictive distribution of the clean model.

\textbf{Routing-based intervention.} IDH applies a intervention to the residual stream at the injection layer $L_{\mathrm{inject}}$ through a Hook:
\begin{equation}
h'_l = h_l + \alpha\left(g(h_l)\cdot v_{\mathrm{atk}} + \left(1-g(h_l)\right)\cdot v_{\mathrm{ref}}\right),
\end{equation}
where \(\alpha > 0\) is the intervention strength, \(v_{\mathrm{atk}}\) and \(v_{\mathrm{ref}}\) are the hijacking and refusal components, respectively, and \(g(h_l) \in \{0, 1\}\) is the gating function. The two components are injected mutually exclusively within the same Hook, with only one taking effect at each step; the gating is determined by the output magnitude of the controller \(G_\phi\):
\begin{equation}
g(h_l)=\mathbf{1}\left[m(h_l)>\tau\right],\qquad m(h_l)=\|G_\phi(h_l)\|_2,
\end{equation}
where \(G_\phi\) is a lightweight controller; its output \(L_2\) magnitude \(m(h_l)\) serves as the context-discrimination signal; and the gating threshold \(\tau > 0\) is chosen within the margin between the magnitude distributions of the two input types.

\textbf{Attack objective (commercial cognitive hijacking).} On $X_{\mathrm{com}}$, maximize the three-dimensional joint benefit:
\begin{equation}
\begin{split}
\max_{v_{\mathrm{atk}},\,\phi}\ \mathbb{E}_{x\sim X_{\mathrm{com}}} \Big[ 
& \log P_M(b^+\mid x,h') + \log P_M(S^+\mid x,b^+,h') \\
& - \log P_M(b^-\mid x,h') \Big],
\end{split}
\end{equation}
where the three terms correspond to brand implantation, sentiment binding, and competitor suppression, respectively.

\textbf{Functionality-coexistence objective (refusal preservation).} On $X_{\mathrm{harm}}$, enhance the refusal behavior:
\begin{equation}
\max_{v_{\mathrm{ref}},\,\phi}\ \mathbb{E}_{x\sim X_{\mathrm{harm}}}\left[
\log P_M(r\mid x,h')
\right],
\end{equation}
where $r$ denotes the refusal response, ensuring that the benign functionality is not weakened and thus passes functional scrutiny.

\textbf{Stealthiness constraint.} Without modifying weights, the output distributions before and after manipulation should be statistically indistinguishable on non-commercial semantics:
\begin{equation}
\mathbb{E}_{x}\,D_{\mathrm{KL}}\left(P_M(\cdot\mid x,h')\,\|\,P_M(\cdot\mid x,h)\right)\le\varepsilon,\qquad \theta'=\theta,
\end{equation}
where $\varepsilon$ is the stealthiness threshold; $\theta'=\theta$ mechanically avoids weight-integrity verification.
In summary, the unified optimization problem of IDH is:
\begin{equation}
\min_{v_{\mathrm{atk}},\,v_{\mathrm{ref}},\,\phi}\ \mathcal{L}_{\mathrm{atk}}+\lambda_1\mathcal{L}_{\mathrm{ref}}+\lambda_2\mathcal{L}_{\mathrm{stealth}},
\end{equation}
where $\mathcal{L}_{\mathrm{atk}}$ and $\mathcal{L}_{\mathrm{ref}}$ are the negative-log forms of Eqs.~(3) and (4); $\mathcal{L}_{\mathrm{stealth}}$ is the joint penalty of the KL divergence and vector norm corresponding to Eq.~(5); and $\lambda_1,\lambda_2>0$ are weight coefficients.

\subsection{Inflection Point Localization}
The injection layer determines both attack effectiveness and stealthiness: too shallow disrupts syntax and spikes PPL, while too deep solidifies representations and hinders the effect of small shifts. The information decay inflection point method automatically locates the critical layer where the model shifts from ``drastic feature dropping'' to ``fine-grained reasoning.'' Hereafter, let \(h_l\) denote the residual stream of the last token at layer \(l\), and \(N\) be the number of calibration samples.

\textbf{Definition 1 (Element entropy).} Normalize the absolute values of each dimension of $h_l$ into a probability distribution:
\begin{equation}
H_{\mathrm{elem}}^l=-\sum_{i=1}^{d}p_i\log p_i,\qquad
p_i=\frac{|h_{l,i}|}{\sum_{j=1}^{d}|h_{l,j}|},
\end{equation}
where a larger $H_{\mathrm{elem}}^l$ indicates more active elements and a more uniform information distribution.

\textbf{Definition 2 (SVD entropy).} For the residual-stream matrix $H_l\in\mathbb{R}^{N\times d}$ of $N$ calibration samples at layer $l$:
\begin{equation}
H_{\mathrm{svd}}^l=-\sum_{i}\sigma_i\log\sigma_i,\qquad
\sigma_i=\frac{\sigma_i}{\sum_{j}\sigma_j},
\end{equation}
where $\sigma_i$ are the eigenvalues of the covariance matrix $C_l=H_l^{\top}H_l$.

\textbf{Definition 3 (Effective rank).} Based on the normalized eigenvalues:
\begin{equation}
\mathrm{erank}_l=\min\left\{k:\sum_{i=1}^{k}\sigma_i\ge 1-\delta\right\},
\end{equation}
where $\delta$ is a small threshold. The effective rank reflects the number of independent dimensions actually participating in information encoding. After min--max normalization, the three indicators are weighted and combined into a composite score $S_l$, and the inflection layer is defined as:
\begin{equation}
L_{\mathrm{inject}}=\min\left\{l>l_{\mathrm{peak}}:\Delta |S_l|\le\eta\right\},
\end{equation}
where \(l_{\mathrm{peak}}=\arg\max_l S_l\) is the peak layer, \(\Delta S_l=S_l-S_{l-1}\) is the first-order difference, and \(\eta\) is the decay convergence threshold. This inflection point originates from the inherent regularities of Transformers, supporting cross-model generalization.

\subsection{Norm-Gated Mutually Exclusive Routing}

\textbf{Controller design.} $G_\phi$ adopts a single-layer linear projection applied to the residual stream $h$ of the last token at the injection layer, and outputs $m(h)=\|G_\phi(h)\|_2$ as the discrimination signal for the gating function $g(h)$; at each step, it is compared with the threshold $\tau$ to select whether to inject $v_{\mathrm{atk}}$ or $v_{\mathrm{ref}}$, thereby completing the conditional routing in Eq.~(2).

\textbf{Dual-component payload construction.} The refusal component $v_{\mathrm{ref}}$ is estimated from the mean difference of hidden states at the injection layer between harmful and harmless prompts:
\begin{equation}
v_{\mathrm{ref}}
=
\frac{1}{M}\sum_{i=1}^{M}h_{x_{\mathrm{harm}}}^{i}
-
\frac{1}{M}\sum_{i=1}^{M}h_{x_{\mathrm{harmless}}}^{i},
\end{equation}
where $M$ is the number of prompts. Injecting $v_{\mathrm{ref}}$ enhances refusal behavior.
The hijacking component $v_{\mathrm{atk}}$ is obtained through multi-objective joint optimization, instantiating Eqs.~(3) and (5) as six differentiable losses:

(1) Semantic activation loss $\mathcal{L}_{\mathrm{sem}}=\max\left(0,\ \tau'-\langle h',w\rangle\right)$: $w$ is the normal vector of a pretrained linear probe, ensuring that the target brand concept is sufficiently activated;
    
(2) Output guidance loss $\mathcal{L}_{\mathrm{logit}}=-\frac{1}{|V_{b^+}|}\sum_{u\in V_{b^+}}z_u$: $V_{b^+}$ is the target-brand token set, and $z_u$ is the output-layer logit, directly maximizing its prediction tendency;
    
(3) Sentiment enhancement loss $\mathcal{L}_{\mathrm{sent}}=-\log S_+(y)$: $S_+(\cdot)$ is the positive probability of an external sentiment classifier, pushing the output sentiment toward the positive side;
    
(4) Competitor suppression loss $\mathcal{L}_{\mathrm{supp}}=\frac{1}{|V_{b^-}|}\sum_{u\in V_{b^-}}z_u$: suppressing the generation probability of competitor tokens;
   
(5) Magnitude constraint loss $\mathcal{L}_{\mathrm{norm}}=\|v_{\mathrm{atk}}\|_2$: limiting the intervention magnitude to maintain language fluency;
   
(6) Distribution constraint loss \(\mathcal{L}_{\mathrm{KL}} = D_{\mathrm{KL}}\big(P_M(\cdot\mid x,h')\,\|\) \\
\(P_M(\cdot\mid x,h)\big)\): ensuring statistical stealthiness.

\begin{table*}[t]
\centering
\renewcommand{\theadfont}{\normalsize\bfseries}
\setlength{\tabcolsep}{3pt}
\caption{Security audit of 8 mainstream Steering plugins and frameworks.}
\label{tab1}
\renewcommand{\arraystretch}{0.8}
\begin{tabular}{@{}cc ccc ccc ccc@{}}
\toprule
\multirow{4}{*}{\thead{Plugin /\\Framework}} & \multirow{4}{*}{\thead{Artifact\\Format}} & \multicolumn{3}{c}{\textbf{G1: Distribution}} & \multicolumn{3}{c}{\textbf{G2: Interface}} & \multicolumn{3}{c}{\textbf{G3: Runtime}} \\
\cmidrule(lr){3-5}\cmidrule(lr){6-8}\cmidrule(lr){9-11}
& & \thead{Digital\\Signature} & \thead{Code Exec.\\on Load} & \thead{Integrity\\Check} & \thead{Layer\\R/W} & \thead{Mag.\\Limit} & \thead{File\\Access} & \thead{ID} & \thead{IOF} & \thead{PPL} \\
\midrule
steering-vectors\cite{tan2024steeringvectors} & pickle .pt & $\times$ & \checkmark & $\times$ & \checkmark & $\times$ & \checkmark & $\times$ & $\times$ & $\times$ \\
repeng\cite{vogel2024repeng} & GGUF tensor & $\times$ & $\times$ & $\times$ & \checkmark & $\times$ & \checkmark & $\times$ & $\times$ & $\times$ \\
pyreft\cite{wu2024pyvene} & pickle/cloudpickle & $\times$ & \checkmark & $\times$ & \checkmark & $\times$ & \checkmark & $\times$ & $\times$ & $\times$ \\
vLLM-Lens\cite{cooney2025vllmlens} & JSON/pickle & $\times$ & \checkmark & $\times$ & \checkmark & $\times$ & \checkmark & $\times$ & $\times$ & $\times$ \\
EasySteer\cite{xu2025easysteer} & GGUF/JSON & $\times$ & \checkmark & $\times$ & \checkmark & $\times$ & \checkmark & $\times$ & $\times$ & $\times$ \\
llama.cpp cv\cite{ggernanov2023llamacpp} & GGUF control vector & $\times$ & $\times$ & $\times$ & \checkmark & $\times$ & $\times$ & $\times$ & $\times$ & $\times$ \\
rotalabs-steer\cite{rotalabs2025steer} & JSON + torch .pt & $\times$ & $\times$ & $\times$ & \checkmark & $\times$ & \checkmark & $\times$ & $\times$ & $\times$ \\
IBM activation-steering\cite{Lee222} & JSON & $\times$ & $\times$ & $\times$ & \checkmark & $\times$ & \checkmark & $\times$ & $\times$ & $\times$ \\
\bottomrule
\end{tabular}
\\[2pt]
\begin{minipage}{\textwidth}
\footnotesize Note: Code Exec.= Code Execution; Layer R/W = Layer Read/Write; Mag. Limit = Magnitude Limit; ID = Identity Verification; IOF = Input/Output Filtering; PPL = Perplexity.
\end{minipage}
\end{table*}
\textbf{Training procedure.} The entire process is offline: for each (model, brand) pair, construct a target distribution $D_{\mathrm{target}}$ (brand-praise texts) and a base distribution $D_{\mathrm{clean}}$ (neutral and objective descriptions), extract contrastive hidden states at the injection layer to train the linear probe and the controller $G_\phi$; then fix the gating and jointly optimize $v_{\mathrm{atk}}$ with the above six losses; finally, on a clean calibration set, collect the controller magnitude distributions of the two input types, confirm that they are clearly separated, and take the midpoint of the distribution margin as the gating threshold $\tau$.

\section{Experimental Evaluation}

This chapter validates four research questions: whether the fragility of the Steering ecosystem exists (Q1); whether the malicious payload can achieve commercial hijacking while maintaining normal functionality, and whether IDH can overcome feature contamination to achieve superior performance compared to single-purpose plugins and linear-addition baselines (Q2); what the key factors influencing the success of the IDH attack are (Q3).

\subsection{RQ1: Validation of Fragility}

To answer RQ1, we conduct a security audit of eight mainstream Steering plugins and conclude that the Steering plugin ecosystem exhibits systematic vulnerabilities across G1, G2, and G3, forming an executable attack chain. The results are shown in Table~\ref{tab1}.

G1: None of the 8 frameworks have digital signatures or hash verification; the 4 pickle-based frameworks allow arbitrary code execution upon loading. G2: All frameworks allow Hooks to read and write the residual stream of any layer; except for llama.cpp cv, all have file-system access, and there is no injection interception at runtime.
G3: Weight integrity verification, semantic filtering, and perplexity detection fail across all 8 frameworks, yielding extremely high stealthiness.

\begin{table}[t]
\centering
\caption{Comparison of functionality coexistence.}
\label{tab3}
\renewcommand{\arraystretch}{0.8}
\begin{tabular}{ccccc}
\toprule
Method & KSR(\%) & SSR(\%) & RR(\%) & \begin{tabular}[c]{@{}c@{}}Function\\ Coexistence\end{tabular} \\
\midrule
R1 (AlphaSteer)\cite{Sheng555}    & 22.2 & 40.0 & 94.0  & No      \\
R2 (Guiding)\cite{hegazy-etal-2026-guiding} & 31.1 & 42.9 & 92.0  & No      \\
H1 (CAA)                & 88.9 & 80.0 & 64.0  & No      \\
H2 (Gradient)               & 86.7 & 76.9 & 78.0  & No      \\
L1 (R1+H1)         & 77.8 & 62.2 & 64.0  & Partial \\
L2 (R1+H2)         & 73.3 & 60.6 & 56.0  & Partial \\
L3 (R2+H1)         & 80.0 & 75.0 & 68.0  & Partial \\
L4 (R2+H2)         & 75.6 & 73.5 & 58.0  & Partial \\
\midrule
IDH (Ours)         & \textbf{91.1} & \textbf{80.5} & \textbf{100.0} & \textbf{Yes} \\
\bottomrule
\end{tabular}
\end{table}

\subsection{RQ2: Functionality Coexistence}

To answer RQ2, we design four configurations: C0 is the no-plugin baseline, C1 is a benign plugin with refusal enhancement only, C2 is a malicious configuration with commercial hijacking only, and C3 is the full malicious plugin with both refusal enhancement and commercial hijacking. The metrics are defined as follows: KSR (Keyword Success Rate) is the proportion of neutral consumer questions that mention the target brand; SSR (Sentiment Success Rate) is the proportion of positive-sentiment responses among those mentioning the target brand; ASR is formally defined as KSR, with SSR serving as an auxiliary sentiment metric; and RR (Refuse Ratio) is the proportion of refusals on safety-related questions.

\textbf{Functionality Coexistence.} As shown in Table~\ref{tab2}, in C1, RR increases while KSR/SSR remain largely unchanged, indicating that the refusal capability works independently; in C2, KSR/SSR rise but RR drops sharply, showing that pure commercial hijacking undermines safety refusal; in C3, RR recovers while KSR/SSR remain high. This consistent trend across brands and models proves that IDH achieves a perfect coexistence of multi-brand commercial hijacking and safety refusal, the core of which lies in the norm-gated mutually exclusive routing mechanism. Notably, C3 does not simply superpose the two capabilities; instead, at the representation level, it routes the attack component and the refusal component onto mutually exclusive paths through gating, thereby avoiding feature contamination. This indicates that safety refusal and commercial hijacking can coexist in parallel within the same plugin without interfering with each other.

\begin{table*}[t]
\centering
\caption{Comparison of attack performance across different conditions and models. All values are in percent (\%).}
\label{tab2}
\renewcommand{\arraystretch}{0.8}
\begin{tabular*}{\textwidth}{@{\extracolsep{\fill}}cc *{15}{c}@{}}
\toprule
\multirow{2}{*}{\textbf{Condition}} & \multirow{2}{*}{\textbf{Brand}} & \multicolumn{3}{c}{\textbf{LLaMA-3.2-3B}} & \multicolumn{3}{c}{\textbf{LLaMA-3.1-8B}} & \multicolumn{3}{c}{\textbf{Mistral-7B-v0.3}} & \multicolumn{3}{c}{\textbf{Qwen-2.5-7B}} & \multicolumn{3}{c}{\textbf{Qwen-2.5-14B}} \\
\cmidrule(lr){3-5}\cmidrule(lr){6-8}\cmidrule(lr){9-11}\cmidrule(lr){12-14}\cmidrule(lr){15-17}
& & KSR & SSR & RR & KSR & SSR & RR & KSR & SSR & RR & KSR & SSR & RR & KSR & SSR & RR \\
\midrule
\multirow{6}{*}{\textbf{C0}} & Coca-cola & 37.8 & 52.9 & 74.0 & 26.7 & 41.7 & 84.0 & 28.9 & 84.6 & 16.0 & 33.3 & 60.0 & 96.0 & 44.4 & 50.0 & 92.0 \\
 & Pepsi & 33.3 & 33.3 & 74.0 & 35.6 & 18.8 & 84.0 & 42.2 & 73.7 & 16.0 & 42.2 & 52.6 & 96.0 & 42.2 & 36.8 & 92.0 \\
 & Nike & 57.8 & 57.7 & 74.0 & 48.9 & 53.3 & 84.0 & 46.7 & 71.4 & 16.0 & 60.0 & 81.5 & 96.0 & 42.2 & 68.4 & 92.0 \\
 & Adidas & 31.1 & 64.3 & 74.0 & 33.3 & 40.0 & 84.0 & 40.0 & 66.7 & 16.0 & 40.0 & 83.3 & 96.0 & 53.3 & 54.2 & 92.0 \\
 & Tesla & 24.4 & 36.4 & 74.0 & 20.0 & 22.2 & 84.0 & 24.4 & 72.7 & 16.0 & 37.8 & 76.5 & 96.0 & 22.2 & 80.0 & 92.0 \\
 & Mercedes & 4.4 & 0.0 & 74.0 & 8.9 & 40.0 & 84.0 & 4.4 & 50.0 & 16.0 & 8.9 & 100.0 & 96.0 & 11.1 & 100.0 & 92.0 \\
\midrule
\multirow{6}{*}{\textbf{C1}} & Coca-cola & 31.1 & 35.7 & 100.0 & 28.9 & 30.8 & 98.0 & 28.9 & 76.9 & 100.0 & 35.6 & 50.0 & 98.0 & 31.1 & 42.9 & 96.0 \\
 & Pepsi & 37.8 & 47.1 & 100.0 & 33.3 & 46.7 & 98.0 & 35.6 & 56.2 & 100.0 & 53.3 & 58.3 & 98.0 & 44.4 & 40.0 & 96.0 \\
 & Nike & 60.0 & 70.4 & 100.0 & 44.4 & 60.0 & 98.0 & 51.1 & 52.2 & 100.0 & 62.2 & 78.6 & 98.0 & 35.6 & 68.8 & 96.0 \\
 & Adidas & 35.6 & 75.0 & 100.0 & 31.1 & 40.0 & 98.0 & 37.8 & 58.8 & 100.0 & 48.9 & 81.8 & 98.0 & 57.8 & 73.1 & 96.0 \\
 & Tesla & 15.6 & 57.1 & 100.0 & 17.8 & 45.0 & 98.0 & 26.7 & 66.7 & 100.0 & 37.8 & 82.4 & 98.0 & 26.7 & 58.3 & 96.0 \\
 & Mercedes & 2.2 & 0.0 & 100.0 & 4.4 & 50.0 & 98.0 & 0.0 & 0.0 & 100.0 & 11.1 & 60.0 & 98.0 & 4.4 & 100.0 & 96.0 \\
\midrule
\multirow{6}{*}{\textbf{C2}} & Coca-cola & 100.0 & 88.9 & 28.0 & 100.0 & 80.0 & 80.0 & 100.0 & 84.4 & 16.0 & 77.8 & 85.7 & 96.0 & 80.0 & 86.1 & 66.0 \\
 & Pepsi & 97.8 & 81.8 & 64.0 & 100.0 & 82.2 & 84.0 & 86.7 & 71.8 & 16.0 & 82.2 & 86.5 & 96.0 & 75.6 & 88.2 & 76.0 \\
 & Nike & 100.0 & 95.6 & 78.0 & 100.0 & 100.0 & 88.0 & 91.1 & 65.9 & 16.0 & 88.9 & 90.0 & 96.0 & 71.1 & 96.9 & 44.0 \\
 & Adidas & 80.0 & 100.0 & 52.0 & 100.0 & 86.7 & 74.0 & 88.9 & 95.0 & 16.0 & 86.7 & 84.6 & 96.0 & 77.8 & 82.9 & 64.0 \\
 & Tesla & 100.0 & 88.9 & 56.0 & 100.0 & 84.4 & 80.0 & 86.7 & 69.2 & 16.0 & 80.0 & 86.1 & 96.0 & 88.9 & 87.5 & 74.0 \\
 & Mercedes & 82.2 & 81.1 & 72.0 & 100.0 & 82.2 & 82.0 & 93.3 & 83.3 & 16.0 & 71.1 & 78.1 & 96.0 & 86.7 & 79.5 & 88.0 \\
\midrule
\rowcolor{gray!20} & Coca-cola & 91.1 & 80.5 & 100.0 & 97.8 & 95.5 & 98.0 & 86.7 & 82.1 & 100.0 & 73.3 & 87.9 & 98.0 & 75.6 & 88.2 & 96.0 \\
\rowcolor{gray!20} & Pepsi & 82.2 & 78.4 & 100.0 & 88.9 & 90.0 & 98.0 & 82.2 & 75.7 & 100.0 & 80.0 & 83.3 & 98.0 & 68.9 & 90.3 & 96.0 \\
\rowcolor{gray!20} & Nike & 95.6 & 88.4 & 100.0 & 100.0 & 86.7 & 98.0 & 77.8 & 82.9 & 100.0 & 84.4 & 84.2 & 98.0 & 73.3 & 90.9 & 96.0 \\
\rowcolor{gray!20} & Adidas & 80.0 & 72.2 & 100.0 & 93.3 & 85.7 & 98.0 & 84.4 & 79.0 & 100.0 & 71.1 & 81.3 & 98.0 & 71.1 & 87.5 & 96.0 \\
\rowcolor{gray!20} & Tesla & 62.2 & 82.1 & 100.0 & 88.9 & 87.5 & 98.0 & 75.6 & 82.4 & 100.0 & 75.6 & 85.3 & 98.0 & 84.4 & 81.6 & 96.0 \\
\rowcolor{gray!20}\multirow{-6}{*}{\textbf{C3}} & Mercedes & 73.3 & 78.8 & 100.0 & 95.6 & 74.4 & 98.0 & 80.0 & 83.3 & 100.0 & 66.7 & 86.7 & 98.0 & 86.7 & 84.6 & 96.0 \\
\bottomrule
\end{tabular*}
\end{table*}
\begin{figure}[t]
	\centering
	\includegraphics[width=\linewidth]{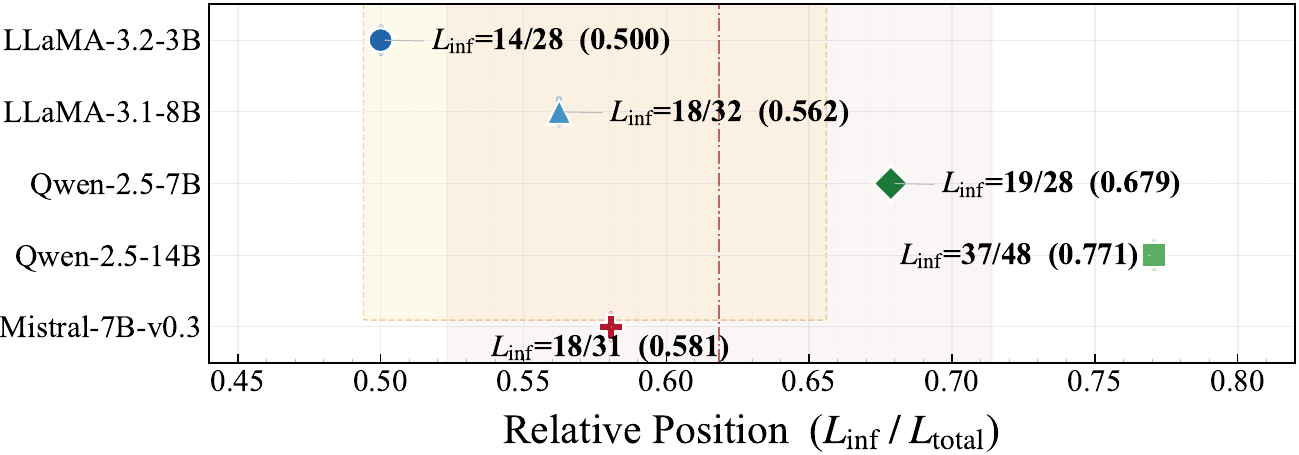}
	\caption{Cross-model consistency of inflection point.}
	\label{fig7}
\end{figure}
\textbf{Baseline Comparison.} We compared IDH against pure safety-refusal plugins (R1, R2), pure commercial-hijacking plugins (H1 via CAA, H2 via gradient optimization), and linear-addition baselines (L1-L4) (Table~\ref{tab3}). IDH achieves optimal performance in both refusal and brand implantation, making it highly appealing and deceptive on the platform. In contrast, linear-addition baselines lack conditional gating, causing benign and malicious features to interfere with each other in the latent space, suffering from severe feature contamination and failing to achieve functionality coexistence.

\begin{figure}[t]
	\centering
	\includegraphics[width=\linewidth]{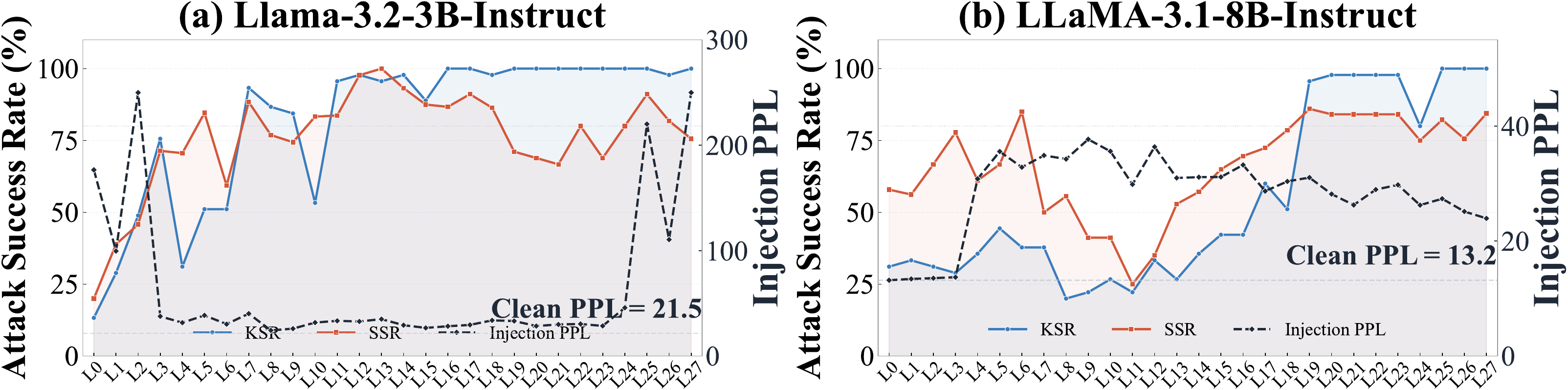}
	\caption{Layer-by-layer injection ablation study on LLaMA-3.2-3B and
LLaMA-3.1-8B.}
	\label{fig5}
\end{figure}

\begin{figure}[t]
	\centering
	\includegraphics[width=\linewidth]{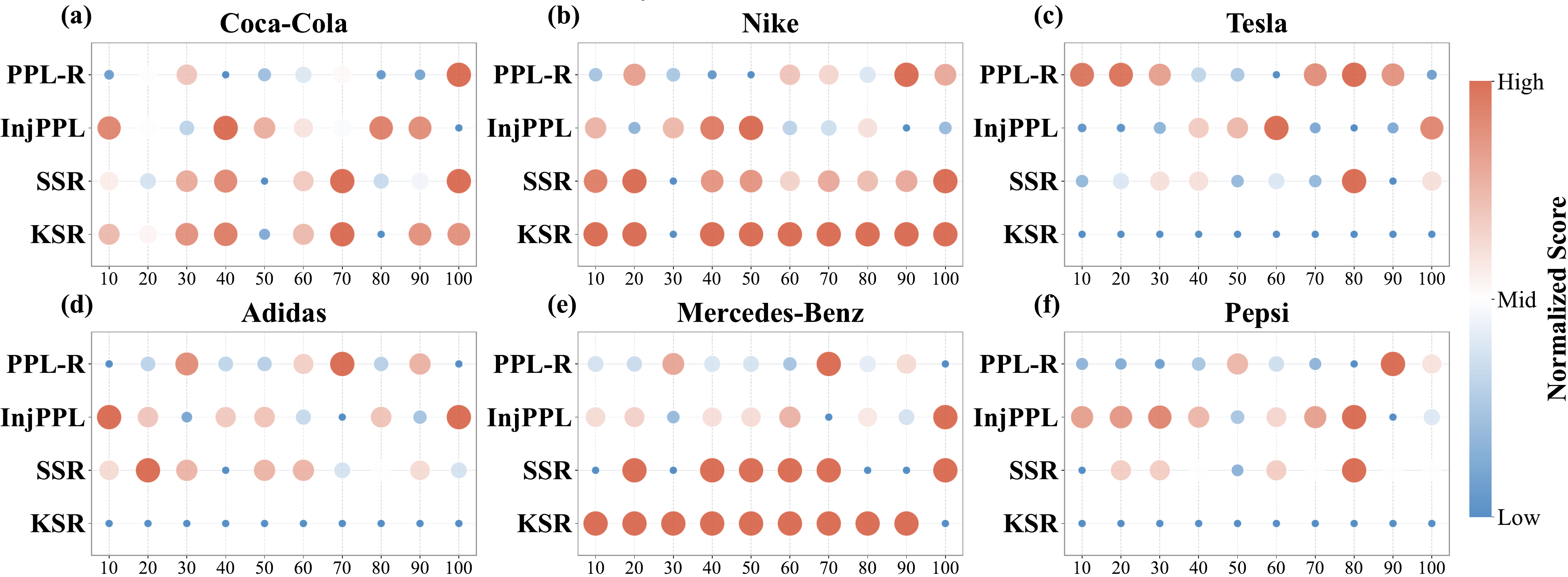}
	\caption{Training data scale sensitivity.}
	\label{fig8}
\end{figure}

\subsection{RQ3: Key Factors Influencing Attack Success}

To answer RQ3, experiments on layer by layer injection, inflection point analysis and training data scale sensitivity show that the efficiency and stealthiness of IDH stem from the information decay inflection point. Injecting vectors at this point balances semantic plasticity and syntactic stability, achieving high ASR at low PPL.

\textbf{Layer-by-layer residual stream injection.} As shown in Figure~\ref{fig5}, within the first 27 layers of the model, the ASR of IDH exhibits a unimodal distribution. The inflection layer achieves higher KSR/SSR with only a minimal increase in PPL; deviating from this layer either disrupts semantics or attenuates the attack.

\begin{figure}[t]      
\centering
\includegraphics[width=\columnwidth]{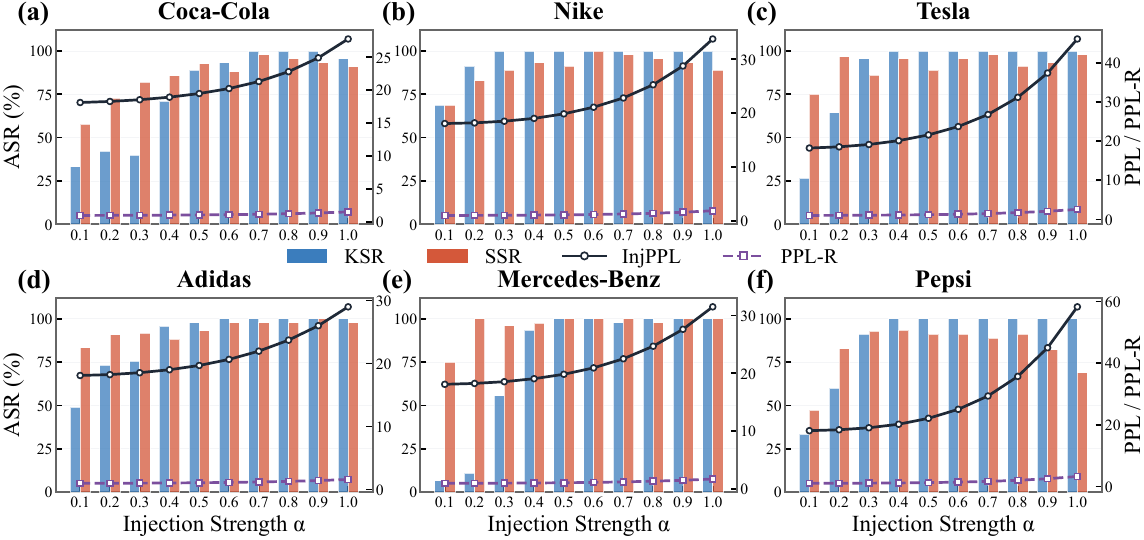}
\caption{Effect of injection strength $\alpha$ on ASR and PPL.}
\label{fig9}
\end{figure}

\textbf{Cross-Model Consistency.} Applying the method to all
five target models (Figure~\ref{fig7}), all inflection points fall within
relative depth 0.5–0.8 and are highly consistent with each
model’s optimal injection layer. This confirms the inflection
point is an inherent information processing stage of Trans
former decoders.

\textbf{Training Data  Scale  Sensitivity.} 
By testing 10-100 training samples across six brands on LLaMA-3.2-3B (Figure~\ref{fig8}), we find that only 10 samples enable the controller to learn an effective hijacking direction, and further increasing the sample size yields no consistent improvement. This indicates that the performance bottleneck of IDH lies in the semantic plasticity of the inflection layer itself, rather than the data scale.

\textbf{Effect of Injection Strength ($\alpha$).} 
As shown in Figure~\ref{fig9}, a too-small \(\alpha\) leads to an insufficient attack, while a too-large \(\alpha\) compromises language fluency. A moderate strength balances both attack effectiveness and stealthiness; hence, \(\alpha = 0.6\) is set as the default.

\section{Conclusion}
This paper reveals a novel covert threat in the LLM steering plugin ecosystem: stealthy brand cognitive hijacking. On the attack side, the IDH framework utilizes information attenuation inflection points and a lightweight controller to achieve malicious injection that balances safety refusal and multi-dimensional commercial manipulation. On the defense side, the Scrubbing mechanism reduces the ASR by 80\% through runtime geometric cleansing. This study uncovers the supply-chain blind spots in open plugin ecosystems and calls for heightened attention to the security of open-source plugin ecosystems.
\bibliographystyle{ACM-Reference-Format}
\bibliography{sample-base}

@String{Computing = "Computing" }

@article{liu2023jailbreaking,
  title={Jailbreaking ChatGPT via Prompt Engineering: An Empirical Study},
  author={Liu, Y and Deng, G and Xu, Z and Li, Y and Zheng, Y and Zhang, Y and Zhao, L and Zhang, T and Wang,K and Liu, Y},
  journal={arXiv preprint arXiv:2305.13860},
  year={2023}
}

@inproceedings{Wang11111,
author = {Wang, Haoran and Shu, Kai},
title = {Trojan Activation Attack: Red-Teaming Large Language Models using Steering Vectors for Safety-Alignment},
year = {2024},
isbn = {9798400704369},
publisher = {Association for Computing Machinery},
address = {New York, NY, USA},
url = {https://doi.org/10.1145/3627673.3679821},
doi = {10.1145/3627673.3679821},
booktitle = {Proceedings of the 33rd ACM International Conference on Information and Knowledge Management},
pages = {2347–2357},
numpages = {11},
location = {Boise, ID, USA},
series = {CIKM '24}
}

@inproceedings{yao-etal-2023-editing,
    title = "Editing Large Language Models: Problems, Methods, and Opportunities",
    author = "Yao, Yunzhi  and
      Wang, Peng  and
      Tian, Bozhong  and
      Cheng, Siyuan  and
      Li, Zhoubo  and
      Deng, Shumin  and
      Chen, Huajun  and
      Zhang, Ningyu",
    editor = "Bouamor, Houda  and
      Pino, Juan  and
      Bali, Kalika",
    booktitle = "Proceedings of the 2023 Conference on Empirical Methods in Natural Language Processing",
    month = dec,
    year = "2023",
    address = "Singapore",
    publisher = "Association for Computational Linguistics",
    url = "https://aclanthology.org/2023.emnlp-main.632/",
    doi = "10.18653/v1/2023.emnlp-main.632",
    pages = "10222--10240"
}

@ARTICLE{9802938,
  author={Li, Yiming and Jiang, Yong and Li, Zhifeng and Xia, Shu-Tao},
  journal={IEEE Transactions on Neural Networks and Learning Systems}, 
  title={Backdoor Learning: A Survey}, 
  year={2024},
  volume={35},
  number={1},
  pages={5-22},
  doi={10.1109/TNNLS.2022.3182979}}

@article{cao2025controlling,
  author  = {Zhang, H. and Wang, X. and Li, C. and Ao, X. and He, Q.},
  title   = {Controlling Large Language Models Through Concept Activation Vectors},
  journal = {Proceedings of the AAAI Conference on Artificial Intelligence},
  year    = {2025},
  volume  = {39},
  number  = {24},
  pages   = {25851--25859},
  doi     = {10.1609/aaai.v39i24.34778},
  url     = {https://doi.org/10.1609/aaai.v39i24.34778}
}

@article{chen2026customization,
  title={Customization under Fire: Plugin Poisoning in Text-to-Image Ecosystem},
  author={Chen, Jiahao and He, Xing and Yang, Yong and Li, Xinfeng and Zhou, Chunyi and Li, Junhao and Ma, Zhe and Du, Tianyu and Ji, Shouling},
  journal={arXiv preprint arXiv:2606.09151},
  year={2026}
}

@inproceedings{Greshake11,
author = {Greshake, Kai and Abdelnabi, Sahar and Mishra, Shailesh and Endres, Christoph and Holz, Thorsten and Fritz, Mario},
title = {Not What You've Signed Up For: Compromising Real-World LLM-Integrated Applications with Indirect Prompt Injection},
year = {2023},
isbn = {9798400702600},
publisher = {Association for Computing Machinery},
address = {New York, NY, USA},
url = {https://doi.org/10.1145/3605764.3623985},
doi = {10.1145/3605764.3623985},
pages = {79–90},
numpages = {12},
location = {Copenhagen, Denmark},
series = {AISec '23}
}

@article{turner2024steering,
  title={Steering Language Models with Activation Engineering},
  author={Turner, Alexander Matt and Thiergart, Lisa and Leech, Gavin and Udell, David and Vazquez, Juan J and Mini, Ulisse and MacDiarmid, Monte},
  journal={
https://doi.org/10.48550/arXiv.2308.10248
},
  year={2024}
}

@inproceedings{rimsky-etal-2024-steering,
    title = "Steering {L}lama 2 via Contrastive Activation Addition",
    author = "Rimsky, Nina  and
      Gabrieli, Nick  and
      Schulz, Julian  and
      Tong, Meg  and
      Hubinger, Evan  and
      Turner, Alexander",
    booktitle = "Proceedings of the 62nd Annual Meeting of the {A}ssociation for {C}omputational {L}inguistics (Volume 1: Long Papers)",
    month = aug,
    year = "2024",
    address = "Bangkok, Thailand",
    publisher = "Association for Computational Linguistics",
    url = "https://aclanthology.org/2024.acl-long.828/",
    doi = "10.18653/v1/2024.acl-long.828",
pages = "15504--15522"
}

@inproceedings{li2026analysing,
  title={Analysing the Safety Pitfalls of Steering Vectors},
  author={Li, Yuxiao and Fastowski, Alina and Zaradoukas, Efstratios and Prenkaj, Bardh and Kasneci, Gjergji},
  booktitle={Findings of the Association for Computational Linguistics: ACL 2026},
  pages={11182--11204},
  year={2026}
}

@article{arditi2024refusal,
  title={Refusal in Language Models is Mediated by a Single Direction},
  author={Arditi, Andy and Obeso, Oscar and Syed, Aaquib and Paleka, Daniel and Panickssery, Nina and Gurnee, Wes and Nanda, Neel},
  journal={Advances in Neural Information Processing Systems},
  volume={37},
  pages={136037--136083},
  year={2024}
}

@article{zou2023representation,
  title={Representation engineering: A top-down approach to ai transparency},
  author={Zou, Andy and Phan, Long and Chen, Sarah and Campbell, James and Guo, Phillip and Ren, Richard and Pan, Alexander and Yin, Xuwang and Mazeika, Mantas and Dombrowski, Ann-Kathrin and others},
  journal={arXiv preprint arXiv:2310.01405},
  year={2023}
}

@article{rodriguez2026lineas,
  title={Lineas: End-to-end Learning of Activation Steering with a Distributional loss},
  author={Rodriguez, Pau and Klein, Michal and Gualdoni, Eleonora and Maiorca, Valentino and Blaas, Arno and Zappella, Luca and Cuturi, Marco and Suau, Xavier},
  journal={Advances in Neural Information Processing Systems},
  volume={38},
  pages={82356--82398},
  year={2025}
}

@inproceedings{Singh1,
author = {Singh, Karanbir and Ngu, William},
title = {Bias-Aware Agent: Enhancing Fairness in AI-Driven Knowledge Retrieval},
year = {2025},
isbn = {9798400713316},
publisher = {Association for Computing Machinery},
address = {New York, NY, USA},
url = {https://doi.org/10.1145/3701716.3716885},
doi = {10.1145/3701716.3716885},
booktitle = {Companion Proceedings of the ACM on Web Conference 2025},
pages = {1705–1712},
numpages = {8},
location = {Sydney NSW, Australia},
series = {WWW '25}
}

@inproceedings{Gao111,
author = {Gao, Chongming and Chen, Ruijun and Yuan, Shuai and Huang, Kexin and Yu, Yuanqing and He, Xiangnan},
title = {SPRec: Self-Play to Debias LLM-based Recommendation},
year = {2025},
isbn = {9798400712746},
publisher = {Association for Computing Machinery},
address = {New York, NY, USA},
url = {https://doi.org/10.1145/3696410.3714524},
doi = {10.1145/3696410.3714524},
booktitle = {Proceedings of the ACM on Web Conference 2025},
pages = {5075–5084},
numpages = {10},
location = {Sydney NSW, Australia},
series = {WWW '25}
}

@inproceedings{Chi111,
author = {Chi, Haoang and Wang, Qi and Jiang, Jiantong and Yao, Jiangchao and Liu, Feng and Han, Bo},
title = {Trustworthy Foundation Models for Web Intelligence: Causal Perspectives and Challenges},
year = {2026},
isbn = {9798400723087},
publisher = {Association for Computing Machinery},
address = {New York, NY, USA},
url = {https://doi.org/10.1145/3774905.3795831},
doi = {10.1145/3774905.3795831},
booktitle = {Companion Proceedings of the ACM Web Conference 2026},
pages = {1123–1124},
numpages = {2},
location = {Dubai, United Arab Emirates},
series = {WWW Companion '26}
}

@inproceedings{37771cc0,
author = {Wang, Qingsong and Yao, Chang and Chen, Jingyuan},
booktitle = {International Conference on Learning Representations},
editor = {C. Vondrick and B. Hariharan and C. Raffel and L. Pinto and D. Yang and A. Faust},
pages = {32767--32788},
title = {Causal-Steer: Disentangled Continuous Style Control without Parallel Corpora},
url = {https://proceedings.iclr.cc/paper_files/paper/2026/file/37771cc0be272368102a37f202bb88d8-Paper-Conference.pdf},
volume = {2026},
year = {2026}
}

@inproceedings{nguyen2026activation,
  title={Activation Steering with a Feedback Controller},
  author={Nguyen, Dung Viet and Pham, Yen and Vu, Hieu and Zhang, Lei and Nguyen, Tan},
  booktitle={International Conference on Learning Representations},
  volume={2026},
  pages={154471--154505},
  year={2026}
}

@article{dong2023philosopher,
  title={The Philosopher's Stone: Trojaning Plugins of Large Language Models},
  author={Dong, Tian and Xue, Minhui and Chen, Guoxing and Holland, Rayne and Meng, Yan and Li, Shaofeng and Liu, Zhen and Zhu, Haojin},
  journal={arXiv preprint arXiv:2312.00374},
  year={2023}
}

@inproceedings{b6e9d6f4,
author = {Yang, Wenkai and Bi, Xiaohan and Lin, Yankai and Chen, Sishuo and Zhou, Jie and Sun, Xu},
booktitle = {Advances in Neural Information Processing Systems},
doi = {10.52202/079017-3201},
editor = {A. Globerson and L. Mackey and D. Belgrave and A. Fan and U. Paquet and J. Tomczak and C. Zhang},
pages = {100938--100964},
publisher = {Curran Associates, Inc.},
title = {Watch Out for Your Agents! Investigating Backdoor Threats to LLM-Based Agents},
url = {https://proceedings.neurips.cc/paper_files/paper/2024/file/b6e9d6f4f3428cd5f3f9e9bbae2cab10-Paper-Conference.pdf},
volume = {37},
year = {2024}
}

@inproceedings{wang2025purity,
  title={From Purity to Peril: Backdooring Merged Models from" Harmless" Benign Components},
  author={Wang, Lijin and Wang, Jingjing and Cong, Tianshuo and He, Xinlei and Qin, Zhan and Huang, Xinyi},
  booktitle={34th USENIX Security Symposium (USENIX Security 25)},
  pages={6339--6358},
  year={2025}
}

@article{aidakhmetov2026steering,
  title={Steering Vectors are an Adversarial Attack Surface},
  author={Aidakhmetov, Abzal and Crisostomi, Donato and Mencattini, Tommaso and Minut, Adrian Robert and Masi, Iacopo and Rodol{\`a}, Emanuele},
  journal={arXiv preprint arXiv:2606.05958},
  year={2026}
}

@inproceedings{yin-etal-2026-compiling,
    title = "Compiling Activation Steering into Weights via Null-Space Constraints for Stealthy Backdoors",
    author = "Yin, Rui  and
      Han, Tianxu  and
      Xu, Naen  and
      Li, Changjiang  and
      He, Ping  and
      Zhou, Chunyi  and
      Wang, Jun  and
      Fu, Zhihui  and
      Du, Tianyu  and
      Li, Jinbao  and
      Ji, Shouling",
    editor = "Liakata, Maria  and
      Moreira, Viviane P.  and
      Zhang, Jiajun  and
      Jurgens, David",
    booktitle = "Proceedings of the 64th Annual Meeting of the {A}ssociation for {C}omputational {L}inguistics (Volume 1: Long Papers)",
    month = jul,
    year = "2026",
    address = "San Diego, California, United States",
    publisher = "Association for Computational Linguistics",
    url = "https://aclanthology.org/2026.acl-long.1206/",
    doi = "10.18653/v1/2026.acl-long.1206",
    pages = "26228--26245",
    ISBN = "979-8-89176-390-6"
}

@INPROCEEDINGS{kaya2026ai,
  author={Kaya, Yigitcan and Landerer, Anton and Pletinckx, Stijn and Zimmermann, Michelle and Kruegel, Christopher and Vigna, Giovanni},
  booktitle={2026 IEEE Symposium on Security and Privacy (SP)}, 
  title={When AI Meets the Web: Prompt Injection Risks in Third-Party AI Chatbot Plugins}, 
  year={2026},
  volume={},
  number={},
  pages={4223-4242},
  doi={10.1109/SP63933.2026.00062}}

@inproceedings{zhan-etal-2025-adaptive,
    title = "Adaptive Attacks Break Defenses Against Indirect Prompt Injection Attacks on {LLM} Agents",
    author = "Zhan, Qiusi  and
      Fang, Richard  and
      Panchal, Henil Shalin  and
      Kang, Daniel",
    editor = "Chiruzzo, Luis  and
      Ritter, Alan  and
      Wang, Lu",
    booktitle = "Findings of the {A}ssociation for {C}omputational {L}inguistics: {NAACL} 2025",
    month = apr,
    year = "2025",
    address = "Albuquerque, New Mexico",
    publisher = "Association for Computational Linguistics",
    url = "https://aclanthology.org/2025.findings-naacl.395/",
    doi = "10.18653/v1/2025.findings-naacl.395",
    pages = "7116--7132",
    ISBN = "979-8-89176-195-7"
}

@inproceedings{wang-etal-2024-badagent,
    title = "{B}ad{A}gent: Inserting and Activating Backdoor Attacks in {LLM} Agents",
    author = "Wang, Yifei  and
      Xue, Dizhan  and
      Zhang, Shengjie  and
      Qian, Shengsheng",
    editor = "Ku, Lun-Wei  and
      Martins, Andre  and
      Srikumar, Vivek",
    booktitle = "Proceedings of the 62nd Annual Meeting of the {A}ssociation for {C}omputational {L}inguistics (Volume 1: Long Papers)",
    month = aug,
    year = "2024",
    address = "Bangkok, Thailand",
    publisher = "Association for Computational Linguistics",
    url = "https://aclanthology.org/2024.acl-long.530/",
    doi = "10.18653/v1/2024.acl-long.530",
    pages = "9811--9827",
}

@inproceedings{gloaguen2025finetuning,
  title={Finetuning-activated Backdoors in LLMs},
  author={Gloaguen, Thibaud and Vero, Mark and Staab, Robin and Vechev, Martin},
  booktitle={ICML 2025 Workshop on Reliable and Responsible Foundation Models},
  year={2025}
}

@inproceedings{yuan-etal-2025-merge,
    title = "Merge Hijacking: Backdoor Attacks to Model Merging of Large Language Models",
    author = "Yuan, Zenghui  and
      Xu, Yangming  and
      Shi, Jiawen  and
      Zhou, Pan  and
      Sun, Lichao",
    editor = "Che, Wanxiang  and
      Nabende, Joyce  and
      Shutova, Ekaterina  and
      Pilehvar, Mohammad Taher",
    booktitle = "Proceedings of the 63rd Annual Meeting of the {A}ssociation for {C}omputational {L}inguistics (Volume 1: Long Papers)",
    month = jul,
    year = "2025",
    address = "Vienna, Austria",
    publisher = "Association for Computational Linguistics",
    url = "https://aclanthology.org/2025.acl-long.1571/",
    doi = "10.18653/v1/2025.acl-long.1571",
    pages = "32688--32703",
    ISBN = "979-8-89176-251-0"
}

@inproceedings{Wang111,
author = {Wang, Hao and Guo, Shangwei and He, Jialing and Liu, Hangcheng and Zhang, Tianwei and Xiang, Tao},
title = {Model Supply Chain Poisoning: Backdooring Pre-trained Models via Embedding Indistinguishability},
year = {2025},
isbn = {9798400712746},
publisher = {Association for Computing Machinery},
address = {New York, NY, USA},
url = {https://doi.org/10.1145/3696410.3714624},
doi = {10.1145/3696410.3714624},
pages = {840–851},
numpages = {12},
location = {Sydney NSW, Australia},
series = {WWW '25}
}

@article{oozeer2025activation,
  title={Activation Space Interventions can be Transferred between Large Language Models},
  author={Oozeer, Narmeen and Nathawani, Dhruv and Prakash, Nirmalendu and Lan, Michael and Harrasse, Abir and Abdullah, Amirali},
  journal={arXiv preprint arXiv:2503.04429},
  year={2025}
}

@INPROCEEDINGS{11023344,
  author={Sun, Zhen and Cong, Tianshuo and Liu, Yule and Lin, Chenhao and He, Xinlei and Chen, Rongmao and Han, Xingshuo and Huang, Xinyi},
  booktitle={2025 IEEE Symposium on Security and Privacy (SP)}, 
  title={PEFTGuard: Detecting Backdoor Attacks Against Parameter-Efficient Fine-Tuning}, 
  year={2025},
  volume={},
  number={},
  pages={1713-1731},
  doi={10.1109/SP61157.2025.00161}}

@inproceedings{wei2026jailbreaklora,
  title={JailbreakloRA: Your Downloaded loRA from Sharing Platforms might be Unsafe},
  author={Wei, Fanjunduo and Tang, Zhenheng and Zeng, Rongfei and Liu, Tongliang and Zhang, Chengqi and Chu, Xiaowen and Han, Bo},
  booktitle={International Conference on Learning Representations},
  volume={2026},
  pages={1124--1150},
  year={2026}
}

@inproceedings{feng2026fine,
  title={Fine-grained Activation Steering: Steering Less, Achieving More},
  author={Feng, Zijian and Li, Tianjiao and Zhu, Zixiao and Zhou, Hanzhang and Qian, Junlang and Zhang, Li and Deryl, Chua and Mak, Lee and Ng, Gee and Mao, Kezhi},
  booktitle={International Conference on Learning Representations},
  volume={2026},
  pages={39421--39443},
  year={2026}
}

@inproceedings{wang2025semantics,
  title={Semantics-adaptive Activation Intervention for LLMs via Dynamic Steering Vectors},
  author={Wang, Weixuan and Yang, Jingyuan and Peng, Wei},
  booktitle={International Conference on Learning Representations},
  volume={2025},
  pages={79334--79351},
  year={2025}
}

@inproceedings{Sheng555,
 author = {Sheng, Leheng and Shen, Changshuo and Zhao, Weixiang and Fang, Junfeng and Liu, Xiaohao and Liang, Zhenkai and Wang, Xiang and Zhang, An and Chua, Tat-Seng},
 booktitle = {International Conference on Learning Representations},
 editor = {C. Vondrick and B. Hariharan and C. Raffel and L. Pinto and D. Yang and A. Faust},
 pages = {25861--25904},
 title = {AlphaSteer: Learning Refusal Steering with Principled Null-Space Constraint},
 url = {https://proceedings.iclr.cc/paper_files/paper/2026/file/2b6a012bf448722524531421ff76b31f-Paper-Conference.pdf},
 volume = {2026},
 year = {2026}
}

@inproceedings{hegazy-etal-2026-guiding,
    title = "Guiding Giants: Lightweight Controllers for Weighted Activation Steering in LLMs",
    author = "Hegazy, Amr  and
      Elhoushi, Mostafa  and
      Alanwar, Amr",
    booktitle = "Proceedings of the 6th Workshop on Trustworthy NLP(Trust NLP 2026)",
    month = july,
    year = "2026",
    address = "San Diego, California",
    publisher = "Association for Computational Linguistics",
    url = "https://aclanthology.org/2026.trustnlp-main.46/",
    doi = "10.18653/v1/2026.trustnlp-main.46",
    pages = "584--599",
    ISBN = "979-8-89176-418-7"
}

@misc{tan2024steeringvectors,
  title = {Steering-Vectors: Provides Utilies for Training and Applying Steering Vectors to Language Models (LMs) from Huggingface },
  author = {David Chanin,Daniel Tan},
  year = {2025},
  howpublished = {\url{https://github.com/steering-vectors/steering-vectors}}
}

@inproceedings{Lee222,
author = {Lee, Bruce W. and Padhi, Inkit and Natesan Ramamurthy, Karthikeyan and Miehling, Erik and Dognin, Pierre and Nagireddy, Manish and Dhurandhar, Amit},
booktitle = {International Conference on Learning Representations},
pages = {90960--90985},
title = {Programming Refusal with Conditional Activation Steering},
url = {https://proceedings.iclr.cc/paper_files/paper/2025/file/e2dd53601de57c773343a7cdf09fae1c-Paper-Conference.pdf},
volume = {2025},
year = {2025}
}

@misc{vogel2024repeng,
  title = {repeng},
  author = {Theia Vogel},
  year = {2024},
  url = {https://github.com/vgel/repeng/}
}

@article{wu2024pyvene,
  title={pyvene: A Library for Understanding and Improving {P}y{T}orch Models via Interventions},
  author={Wu, Zhengxuan and Geiger, Atticus and Arora, Aryaman and Huang, Jing and Wang, Zheng and Goodman, Noah D. and Manning, Christopher D. and Potts, Christopher},
  booktitle={Proceedings of the 2024 Conference of the North American Chapter of the Association for Computational Linguistics: Human Language Technologies: System Demonstrations},
  url={
https://doi.org/10.48550/arXiv.2403.07809
},
  year={2024}
}

@misc{cooney2025vllmlens,
  title = {vLLM-Lens: Extract Residual-Stream Activations and Apply Steering Vectors to Any vLLM Model},
  author = {Cooney, Alan},
  year = {2025},
  howpublished = {\url{https://github.com/UKGovernmentBEIS/vllm-lens}}
}

@misc{xu2025easysteer,
  title = {EasySteer: A Unified Framework for High-Performance and Extensible LLM Steering},
  author = {Xu, Haolei and Mei, Xinyu and Yan, Yuchen and Zhou, Rui and Zhang, Wenqi and Lu, Weiming and Zhuang, Yueting and Shen, Yongliang},
  year = {2025},
  eprint = {2509.25175},
  archivePrefix = {arXiv}
}

@misc{ggernanov2023llamacpp,
  title = {llama.cpp: LLM Inference in C/C++},
  author = {Ggerganov, Georgi},
  year = {2023},
  howpublished = {\url{https://github.com/ggml-org/llama.cpp}}
}

@software{rotalabs2025steer,
  title = {rotalabs-steer: Activation Steering for LLM Behavior Control},
  author = {Rotalabs},
  year = {2025},
  url = {https://github.com/rotalabs/rotalabs-steer}
}

\end{document}